\RequirePackage{fix-cm}
\documentclass[twocolumn, natbib, final]{svjour3}          
\smartqed  

\usepackage{colortbl}
\usepackage[ruled,vlined,titlenumbered]{algorithm2e}
\usepackage[svgnames]{xcolor}
\usepackage{array,multirow}
\usepackage{footnote}
\usepackage {longtable}
\usepackage[counterclockwise]{rotating}
\usepackage{rotate}
\usepackage{rotating}
\usepackage{graphicx}
\usepackage{caption}
\usepackage[flushleft]{threeparttable}
 \usepackage{mathptmx}      
 \journalname{Journal of Ambient Intelligence and Humanized Computing}
\begin{document}

\title{Hybrid Microaggregation for Privacy-Preserving Data Mining
}


\author{Balkis Abidi   \and
       Sadok Ben Yahia \and 
       Charith Perera 
}


\institute{Balkis Abidi \at
              LIPAH, Faculty of Sciences of Tunis, University of El-Manar, Tunisia \\
           \and
           Sadok Ben Yahia \at
              LIPAH, Faculty of Sciences of Tunis, University of El-Manar, Tunisia \\
              \and 
              Charith Perera \at
              School of Computing Science, Newcastle University, United Kingdom\\
}

\date{Received: date / Accepted: date}

\maketitle

\begin{abstract}
$k$-Anonymity by microaggregation is one of the most commonly used anonymization techniques. This success is owe to the achievement of a worth of interest trade-off between information loss and identity disclosure risk. 
However, this method may have some drawbacks. On the disclosure limitation side, there is a lack of protection against attribute disclosure. On the data utility side, dealing with a real datasets is a challenging task to achieve. Indeed, the latter are characterized by their large number of attributes and the presence of noisy data, such that outliers or, even, data with missing values. Generating an anonymous individual data useful for data mining tasks, while decreasing the influence of noisy data is a compelling task to achieve.\\
In this paper, we introduce a new microaggregation method,  called \textsc{HM-pfsom}, based on fuzzy possibilistic clustering. Our proposed method operates through an hybrid manner. This means that the anonymization process is applied per block of similar data. Thus, we can help to decrease the information loss during the anonymization process. 
The \textsc{HM-pfsom} approach proposes to study the distribution of confidential attributes within each sub-dataset. Then, according to the latter distribution, the privacy parameter $k$ is determined, in such a way to preserve the diversity of confidential attributes within the anonymized microdata. This allows to decrease the disclosure risk of confidential information.

\keywords{Hybrid micoaggregation \and Information loss \and Identity disclosure risk \and Attribute disclosure risk \and Fuzzy possibilistic clustering }
\end{abstract}

\section{Introduction}
\label{intro}
The ever growing privacy concern has been a major obstacle for individual data analysis. In fact, many situations require that governmental, public and private organizations share and release their specific data \citep{Chui2014}. These latter, generally, reflect our everyday life activities, \textit{e.g.} credit card transactions, activities on the web, phone calls, widespread diseases, \textit{etc}. Publishing and releasing such type of data can provide benefits to researchers and decision makers, owe to their flexibility and availability of detailed information \citep{Teplitzky2017}. For example, healthcare organizations collect and analyze medical data for the discovery of new drugs and therapies \citep{Rider2013}. Retail companies need information about customers, in order to identify customer purchases behaviours, discover customer shopping patterns and trends, and thus improve the quality of customer services \citep{Greet2014}. Banks and financial institutions also collect financial data to predict credit fraud, evaluate risk and perform trend analysis \citep{Bennardo2015}. As for telecommunication companies, they maintain a great deal of call detail data, which describe the calls that traverse telecommunication networks \citep{Chittaranjan2013}. Such data can be useful to identify vulnerabilities of networks. Social networks are undoubtedly the most extreme example of data valorisation. The deal is to provide users a free social media platform to entertain, in return collect all kinds of information describing the users' relationships, interests, apps in use, and also religion or political opinions \citep{Johnson2012}. Such information are used to sell advertising and insights based on their profiles. The Facebook-Cambridge Analytica data scandal is the prime example of how personal data could be disclosed, where  the collection of personally identifiable information of up to 87 million Facebook users was allegedly used to attempt to influence voter opinion on behalf of politicians who hired them \citep{Solon2018}.

However, individual data may contain confidential information. Thus, collecting, analyzing and sharing such data, raises threat to individual privacy. Data de-identification, \textit{i.e.} hiding explicit \textit{identifiers}, is considered of paramount importance to avoid sensitive information from being disclosed. Such process involves removing any information which is able to \emph{uniquely} identify an individual, \textit{e.g.} name, SSN, \textit{etc} \citep{Garfinkel2015}. Nevertheless, the latter process could not guarantee efficient security. Indeed, it was shown that is possible to manipulate de-identified datasets and recover personal information, through data linkage techniques \citep{ohm2010}.
It is worth mentioning that disclosure risk can be classified into two categories, namely \citep{Hundepool2012}: 
\begin{itemize}
\item \textit{Identity disclosure}: The intruder is able to determine the real identity of individual corresponding to a record in the published microdata. Thus, the intruder can associate the confidential information to the re-identified data subject.
\item \textit{Attribute disclosure}: Even if identity disclosure does not happen, it may be possible for an intruder to infer some information for a specific individual based
on the published microdata.
\end{itemize}

Therefore, a large number of Privacy-Preserving Data Mining (PPDM) methods have been proposed aiming at ensuring privacy of the respondents, while preserving the statistical utility of the original data \citep{Aggarwal2008}. The basic idea of this research area is to modify the collected data, subject to be released, in such a way to perform effectively analyses and knowledge discovery tasks without compromising the security of sensitive information contained in the data. Such process leads to reduce the granularity of information, which can cause a loss of data effectiveness. Thus, finding a trade-off between the two conflicting principles, \textit{i.e.} privacy and data utility, is of the utmost importance in PPDM process. Microaggregation \citep{Ferrer08}, is a widely accepted PPDM method for data anonymization. The principle is to de-associate the relationship between the identity of data subjects and their confidential information. Given a security parameter $k$, the basic idea of microaggregation is to split a dataset into small \textit{groups}, of size at least $k$. Then, the values of the original data are replaced by those of the cluster's centroid to which they belong to. Thereby, any data is indistinguishable among other $(k-1)$ data. The resulting anonymous dataset fulfils the \textit{k-anonymity} model \citep{Sweeney2002}. Thereby, privacy is ensured by preventing record linkage. Meanwhile, data utility is maintained by gathering records that share the same characteristics. However, generating an \textit{optimal} $k$-partition while maintaining the homogeneity of data within a fixed size group has been shown a NP-hard problem \citep{Oganian01}. So, the only practical microaggregation methods are based on heuristics. Generally, these methods rely on \textit{refinement} steps, during the partitioning process, which consists in merging or splitting the obtained fixed size groups. However, in real datasets the poor homogeneity within the generated partition can be significant. Thus, its refinement could be costly and does not necessarily converge to the optimal partition. Besides, all microaggregation methods have focused on decreasing the information loss, while maintaining the constraint of the group's size. As consequence, the modified dataset can be exposed to attribute linkage.  In fact, $k$-Anonymity  can  create  groups  that  leak
information due to lack of diversity in the sensitive attribute \citep{Machanavajjhala2007}.

This study aims to deal with these shortcomings, by proposing a new algorithm called \textsc{HM-pfsom}. The proposed algorithm relies on the following assumptions:
\begin{itemize}
\item Preventing the identity disclosure risk by requiring that the generated partition should fulfil the k-anonymity property.
\item Preventing the attribute disclosure by ensuring the diversity of confidential attributes within each fixed size group of the generated partition.
\item Maintaining the data utility of the anonymous microdata by increasing the homogeneity within the fixed size groups, \textit{i.e.} $k$-partition.
\end{itemize}
In order to meet the latter requirements, the \textsc{HM-pfsom} algorithm operates in an \textit{hybrid} manner, \textit{i.e.} per block of data. Indeed, to avoid the risk of gathering data with dissimilar quasi-identifiers in a same group, the \textsc{HM-pfsom} algorithm splits the microdata into disjoint sub-microdata, through a fuzzy clustering algorithm. The clustering process is applied according to the quasi-identifiers, in order to apply the microaggregation process independently on each sub-microdata, \textit{i.e.} group of data having similar quasi-identifiers.\\ 
Unlike the standard microaggregation methods, the \textsc{HM-pfsom} algorithm don't require a \textit{predefined} privacy parameter $k$, fixed arbitrary, to build the $k$-anonymous microdata. It proposes to study the distribution of confidential attributes within each sub-microdata. Then, according to the latter distribution, the parameter $k$ is determined, in such a way to preserve the diversity of confidential attributes within the anonymized microdata.\\

The paper is organized as follows. Section \ref{section2} discusses the main PPDM approaches, in particular microaggregation methods. Section \ref{section3} introduces the novel \textbf{\textsc{HM-pfsom}} algorithm for \textbf{H}ybrid \textbf{M}icroaggregation by using the \textbf{\textsc{pfsom}} clustering method. Finally, section \ref{section4} evaluates the performance of the proposed approach, through experiments carried out on real-life datasets. Finally, Section \ref{section5} sketches our contributions and points out avenues of future work

\section{Related work and motivation}
\label{section2}
In this section we present a succinct introduction to data anonymization and sanitization techniques. We focus in particular on microaggregation model, which is one of the most popular, studied and used PPDM methods. A discussion is also presented, in order to
situate our contribution with respect to the reviewed ones of the literature.
\subsection{Data anonymization and sanitization techniques}
Let $X = \lbrace x_{1}, x_{2}, \ldots, x_{n} \rbrace$ be a de-identified dataset, also called \textit{microdata}, subject to be released. Each input data $ x_{i} \in X $ is a set of $m$ attributes $ A = \lbrace a_{1},  a_{2}, \ldots, a_{m} \rbrace$, and $x_{i} \left[  a_{j} \right]  $ denotes the value of attribute $a_{j}$ of the data $x_{i}$. We assume that $X$ is a subset of some larger population $\Omega$ where each input data represents an individual. The set of attributes $A$ is composed primarily by \citep{Sweeney2002}: 
\begin{itemize}
\item \textit{Quasi-identifiers} $Qid = \lbrace qid_{1}, qid_{2}, \ldots, qid_{p}\rbrace$, which are a non-sensitive attributes. Nevertheless, if the latter are linked with external information they can \textit{uniquely} identify at least one individual. 
\item \textit{Confidential attributes} $S = \lbrace s_{1}, s_{2}, \ldots, s_{q}\rbrace$, also called \textit{sensitive attributes}, which their values for any particular individual must be kept secret from users who have no direct access to the original data; 
\end{itemize}
Thus, the set of attributes $A$ can be represented by $Qid \cup S$, \textit{i.e.}
$A  = \lbrace qid_{1}, qid_{2}, \ldots, qid_{p}\rbrace \cup \lbrace s_{1}, s_{2}, \ldots, s_{q}\rbrace$, where $p + q = m$.\\

Perturbation of data is a  very easy and effective method for protecting the sensitive information of individual data from unauthorised users or hackers. These methods allow the release of the entire microdata, although by masking the sensitive information \citep{LiuGianKar2008}. This requires to generate a set of random values, which will be used subsequently to hide sensitive information of the released data \citep{Kargupta2005}.
To maintain privacy and data utility, data perturbation methods aim to replace the  original  values of data  with some  artificial  values, while maintaining their statistical properties. As  the  perturbed  data  records  does  not  match  with  the  original  ones,  the  attacker  cannot  recover  the  sensitive information from the perturbed data. \\
The most widely used techniques, in data perturbation methods, are additive noise and multiplicative stochastic noise \citep{Agrawal2000}\citep{Chen2008}\citep{Ciriani2007} \citep{Du2003}\citep{Evfimievski2002}\citep{Mivule2013}.
To maintain the effectiveness of data, the generated random values require to comply with the distribution of the original ones. Thus, the distribution of  each  data  dimension  is  reconstructed  independently. This means that any distribution based data  mining  algorithm  works  under  an  implicit  assumption  to  treat  each  dimension  
independently \citep{Matwin2013}. However, in many cases, a lot of relevant information for data mining algorithms is hidden in inter-attribute correlations. \\
Microaggregation technique \citep{Ferrer08} relaxes the constraint of hiding sensitive attributes with random values. Its principle consists in de-associating the relationship between quasi-identifiers and confidential attributes of individual records. Given a security parameter $k$, the basic idea of microaggregation is to split a dataset into small \textit{groups}, of size at least $k$. Then, the quasi-identifiers of the original data are replaced with those of the cluster's centroid to which they belongs to. Thereby, any data is indistinguishable among other $(k-1)$ data. The resulting anonymous dataset fulfils the \textit{k-anonymity} model \citep{Sweeney2002}. Thus, privacy is ensured by preventing record linkage. Meanwhile, data utility is maintained by gathering records sharing the same characteristics of quasi-identifiers. 
\subsection{The k-anonymous microaggregation model}
Normally, microaggregation gathers the closest data in the same fixed size group, in such a way that the respective distances between the data vectors and the corresponding centroids is as small as possible. Thus, the microaggregation generates a protected microdata $X'$ that is similar to the original one, but where data in $X'$ are slightly different from those of $X$. The optimal $k$-partition is defined as that it maximizes the within-group homogeneity. The higher the within-group homogeneity, the lower the information loss is. However, finding the optimal cluster configuration has been shown to be a NP-hard problem \citep{Oganian01}. This issue has grasped the interest of the literature and a wealthy number of methods exist.\\
According to data dimensionality, microaggregation methods can be split into two categories, namely, \textit{univariate} and  \textit{multivariate} approaches. The $k$-partitioning mechanism is the same in all such methods: first, data vectors are sorted in ascending or descending order according to some criterion. Then, groups of successive $k$ vectors are combined. Inside each group, the effect for each variable is to replace the $k$ values taken by the variable with their average. If the total number of data vectors $n$ is not a multiple of $k$, the last group will contain more than $k$ data vectors.

\subsubsection{Univariate microaggregation methods}
Univariate microaggregation performs a straightforward \textit{one-dimensional} sorting. To this end, projected methods, also called \textit{single axis}, are used to summarize the $p$ quasi-identifiers of each data vector into a \textit{single} value \citep{analysisDR}. The most commonly used methods are \textit{Principal Components Analysis} and the sum of \textit{Zscores} \citep{DRmultiMicro}\citep{Rpackage}. To do so, all attributes are firstly standardized and, for each data vector, the standardized values are added. Vectors are subsequently sorted, \textit{w.r.t.} the scores of the first principal component or by their sum of z-scores. This approach has been shown to be very useful with highly correlated data, \textit{i.e.}, the higher the correlation is, the lower the information loss \citep{DRmultiMicro}. However, in real-life datasets, data are not necessarily so highly correlated, which makes these approaches so ineffective.\\
Another alternative for univariate microaggregation is to apply an anonymization process to each variable independently, \textit{i.e.} data vectors are sorted by the first variable, then groups of $k$ successive values of the first variable are formed and, inside each group, values are replaced by the group average. A similar procedure is repeated for the remainder of variables. This method is referred to as \textit{individual sorting} \citep{mdav1}. Even though, this method usually maintains the data utility, it has a higher disclosure risk, \textit{i.e.} no $k$-anonymity will be achieved in general. Indeed, by just taking into account the first variable, the $k$ data standing in the same cluster might be assigned to different clusters when all the other variables are considered. 
\subsubsection{Multivariate microaggregation methods}
When the multivariate data are microaggregated without projecting them in one-dimension, this is referred to as a \textit{multivariate microaggregation}. The Maximum Distance to Average Vector (MDAV) algorithm \citep{mdav2} is the most used for microaggregartion, which operates through an iterative process. Its principle involves computing the centroid of the quasi-identifiers of all input data. The distance of the data to the obtained centroid is used as a sorting criteria. To achieve that, two  extreme data, $x_{r}$ and $x_{d}$, relative to the centroid are extracted. Where $x_{r}$ is the most distant data vector to the centroid, and $x_{d}$ is the most distant data vector to $x_{r}$. Then, two groups are formed with size $k$ around $x_{r}$ and $x_{d}$, respectively. Such process is repeated until all input data vectors of the original microdata are partitioned. Then, the MDAV algorithm generates a partition of fixed size groups having a same cardinality, which is equal to $k$. Note that, if the number of input data is not divisible by $k$, the cardinality of one group, generally the last one, ranges between $k$ and $2k-1$. However, the obtained partition of the MDAV algorithm may lack flexibility for adapting the group size constraint to the distribution of the data vectors \citep{vmdav1}.
Several microaggregation methods have been proposed to improve the homogeneity within the fixed size groups obtained by the MDAV algorithm  \citep{TFRP} \citep{vmdav1} \citep{Ferrer2010}  \citep{DensityMicro} \citep{BallesteSDM07} \citep{SolanasGM12}. Generally these methods add further steps, called \textit{refinement} steps, which consists in merging or splitting the obtained fixed size groups. Thus, yielding a more freedom partition, having  groups with different cardinality varying between $k$ and $2k -1$. \\
However, in real datasets the poor homogeneity within the generated partition can be significant. So, its refinement could be costly and does not necessarily converge to the optimal partition. Besides, all microaggregation methods have focused on the constraint of homogeneity of the $k$-partition to reduce the information loss. As consequence, the modified dataset can be exposed to an attribute disclosure risk. 
\subsection{Discussion \& Motivation}
In our opinion, the major weakness of micoaggregation methods consists in applying the $k$-partitioning process without studying the distribution of the input data and their correlation. We are convinced that, if there is a step to add, in order to converge to the optimal $k$-partition, it should be applied \emph{before} the partitioning process. This step should analyze the similarity between the quasi-identifiers of the input data, in order to decide which data should be gathered in a same group. Therewith, we are of the view that to prevent against identity disclosure, the parameter $k$ should be fixed according to the distribution of sensitive information, in order to ensure that the $k$-anonymous records fulfil a \textit{diversity} of sensitive information and so prevent attribute disclosure. \\
To meet these issues, we propose a new method, called \textsc{HM-pfsom}, for hybrid microaggregation, by using \textsc{pfsom} algorithm for fuzzy possibilistic clustering \citep{Abidi2013}. The main idea of the  \textsc{HM-pfsom} algorithm consists in splitting the original dataset into disjoint sub-datasets, in such a way that data within the same sub-dataset must be similar to some extend. In addition, they should be dissimilar to those data in other sub-datasets. Such process enables to avoid the refinement phases used to fine tune the $k$-anonymous partition, since the partitioning process is performed only on similar data. Indeed, the \textsc{HM-pfsom} algorithm applies the $k$-partitioning process independently on each sub-dataset.\\
On privacy side, the \textsc{HM-pfsom} algorithm proposes to study the distribution of confidential attributes within each sub-dataset. Then, according to the latter distribution, the privacy parameter $k$ is determined, in such a way to maintain the diversity of confidential attributes within the anonymous microdata.

\section{Data anonymisation by hybrid microaggregation}
\label{section3}
In the following, we introduce a new method, called \textsc{HM-pfsom}, to balance between the above conflicting issues, in order to achieve the tedious optimal partition used to generate an anonymous microdata. 
\subsection{General principle of the hybrid microaggregation}
We propose a new microaggregation method, called hybrid microaggregation, for privacy-preserving data mining. The proposed method, sketched in Algorithm
\ref{architectureMicroFSOM}, follows the following steps:
\begin{enumerate}
\item \textit{Splitting step}: This step aims to split the original microdata into disjoint sub-microdata.
\item \textit{Partitioning step}: This step generates a fixed size partition from each sub-microdata.
\item \textit{Merging step}: In this step, the generated fixed size partitions are used to train a $k$-anonymous microdata.
\end{enumerate}
\linesnumbered 
\begin{algorithm}[htbp]
 \SetVline
 \setnlskip{-3pt}
 \Entree {
$X$ : The original microdata   
  }
 \Sortie{
  $X'$ : The anonymous microdata
  }
  \Deb{
          Split the microdata $X$ into $c$ disjoint sub-microdata\\
           $X = Xid_{1} \cup Xid_{2} \cup \ldots \cup Xid_{c} $.\\
          \ForEach{sub-microdata $Xid_{j} \in \lbrace Xid_{1}, Xid_{2}, \ldots, Xid_{c}  \rbrace$, $\forall j \in \lbrace 1, \ldots, c \rbrace$}{
          	$X'_{j}$ $\leftarrow$  Partitioning\_process($Xid_{j}$)\\
          } 
          $X' = X'_{1} \cup X'_{2} \cup \ldots \cup   X'_{c} $\\
  }
\caption{The general principle of the hybrid microaggregation method}
\label{architectureMicroFSOM}
\end{algorithm}
The hybrid microaggregation integrates an additional step which consists in discovering the distribution of the quasi-identifiers of all input data. This step aims to decide which data that \textit{can not} belong to a same fixed-size group. Doing so, the hybrid microaggregation starts by splitting the original microdata $X$ into a set of disjoint sub-microdata, \textit{i.e.} $X = \lbrace X_{1}, X_{2}, \ldots, X_{c}\rbrace$, where each sub-microdata gathers similar data. Then, an adaptive partitioning process is applied independently on each sub-microdata $X_{i}, i = \lbrace 1, \ldots, c\rbrace$, to train a $k_{i}$-partition. Afterward, the anonymous microdata is obtained from the generated $k_{i}$-partition, $ i = \lbrace 1, \ldots, c\rbrace$. \\
The adaptive partitioning process should maintain the diversity of confidential attributes within each fixed size group. The latter constraint is paramount importance  in the sake of avoiding attribute disclosure risk. To better understand such constraint, let $X_{micro}$, given in Table \ref{Xmicro2}, be an original microdata. Each input data $x_{i} \in X_{micro}$ is characterized by a set of two-dimensional quasi-identifiers, \textit{i.e.} \emph{ZIP code} and \emph{Age}, and one confidential attribute, \textit{i.e.} \emph{Disease}. \\
\begin{table}[htbp]
\caption{A microdata sample $X_{micro}$ }
\label{Xmicro2}
\centering
\begin{tabular}{lccl}
\hline\noalign{\smallskip}
& ZIP code & Age & Disease\\
\noalign{\smallskip}\hline\noalign{\smallskip}
$x_{1}$ & $2025$ & $28$ & Heart Disease \\
$x_{2}$ &$ 2022$ & $29$ & Heart Disease \\
$x_{3}$ & $2022$ & $25$ & Viral Infection\\
$x_{4}$ & $2020$ & $24$ & Viral Infection\\
$x_{5}$ & $1012$ & $50$ & Cancer\\
$x_{6}$ & $1012$ & $55$ & Heart Disease\\
$x_{7}$ & $1013$ & $47$ & Viral Infection\\
$x_{8}$ &$1013$ & $49$ & Viral Infection\\
$x_{9}$ & $1023$& $31$ & Cancer\\
$x_{10}$ & $1022$ & $34$ & Cancer \\
$x_{11}$& $1021$ & $35$ & Cancer \\
$x_{12}$  &$1021$ & $37$ & Cancer \\
\noalign{\smallskip}\hline
\end{tabular}
\end{table}
\begin{table}[h]
\caption{The $4$-anonymous microdata of $X_{micro}$}
\label{Xmicro2anonym}
\centering
\begin{tabular}{clccl}
\hline\noalign{\smallskip}
 & & ZIP code & Age & Disease\\
\noalign{\smallskip}\hline\noalign{\smallskip}
\multirow{4}{*}{$ G_{1} $} & $x_{1}$ & $\textbf{2022}$ & $\textbf{27}$ & Heart Disease \\
& $x_{2}$ &$\textbf{2022}$ & $\textbf{27}$ & Heart Disease \\
& $x_{3}$ & $\textbf{2022}$ & $\textbf{27}$ & Viral Infection\\
& $x_{4}$ & $\textbf{2022}$ & $\textbf{27}$ & Viral Infection\\
\noalign{\smallskip}\hline\noalign{\smallskip}
\multirow{4}{*}{$ G_{2} $} & $x_{5}$ & $\textbf{1012}$ & $\textbf{50}$ & Cancer\\
& $x_{6}$ & $\textbf{1012}$ & $\textbf{50}$ & Heart Disease\\
& $x_{7}$ & $\textbf{1012}$ & $\textbf{50}$ & Viral Infection\\
& $x_{8}$ &$\textbf{1012}$ & $\textbf{50}$ & Viral Infection\\
\noalign{\smallskip}\hline\noalign{\smallskip}
\multirow{4}{*}{$ G_{3} $}& $x_{9}$ & $\textbf{1021}$ & $\textbf{34}$ & \textcolor{red}{Cancer}\\
& $x_{10}$ & $\textbf{1021}$ & $\textbf{34}$ & \textcolor{red}{Cancer} \\
& $x_{11}$& $\textbf{1021}$ & $\textbf{34}$ &\textcolor{red}{Cancer}\\
& $x_{12}$  &$\textbf{1021}$ & $\textbf{34}$ & \textcolor{red}{Cancer} \\
\noalign{\smallskip}\hline
\end{tabular}
\end{table}
Table \ref{Xmicro2anonym} shows an example of a $k$-anonymous microdata generated from the original microdata $X_{micro}$, via a standard microaggregation technique, where the privacy parameter $k$ is set to $3$. 
Note that, the microaggregation is applied in such a way that the obtained anonymous microdata fulfils the $k$-anonymity property. For any combination of values of quasi-identifiers in the released microdata, there are at least $3$ records sharing that combination of values. The fixed size partition is obtained such that each $3$ similar data are gathered in same group. The similarity is measured on the basis of the quasi-identifiers attributes. \\
However, an anonymous microdata, as given in Table \ref{Xmicro2anonym}, can be sensitive to attribute linkage attack, \textit{i.e.} an intruder can discover the confidential information of a given individual. Namely, suppose that the intruder has some knowledge about an individual, for example he is aged $35$ years old and he lives in a city where its \emph{ZIP code} is $1022$. Thus, by linking these information with the $3$-anonymous microdata, the intruder can conclude that the individual in question corresponds to one of the four records having a cancer disease.\\
To prevent such disclosure, it would be better to ensure, during the partitioning process, the \textit{diversity} of confidential attributes within each fixed size group, as shown in Table \ref{opXmicro2partition}.\\
\begin{table}[htbp]
\caption{The optimal $3$-partition of $X_{micro}$}
\label{opXmicro2partition}
\centering
\begin{tabular}{clccl}
\hline\noalign{\smallskip}
 & & ZIP code & Age & Disease\\
\noalign{\smallskip}\hline\noalign{\smallskip}
\multirow{3}{*}{$ G_{1} $}  & $x_{6}$ & $1012$ & $55$ & \textcolor{blue}{Heart Disease}\\
& $x_{5}$ & $1012$ & $50$ & \textcolor{red}{Cancer}\\
& $x_{8}$ &$1013$ & $49$ & \textcolor{magenta}{Viral Infection}\\
\noalign{\smallskip}\hline\noalign{\smallskip}
\multirow{4}{*}{$ G_{2} $} & $x_{7}$ & $1013$ & $47$ & \textcolor{magenta}{Viral Infection}\\
& $x_{12}$  &$1021$ & $37$ & \textcolor{red}{Cancer} \\
& $x_{10}$ & $1022$ & $34$ & \textcolor{red}{Cancer} \\
& $x_{2}$ &$ 2022$ & $29$ & \textcolor{blue}{Heart Disease} \\
\noalign{\smallskip}\hline\noalign{\smallskip}
\multirow{5}{*}{$ G_{3} $} & $x_{11}$& $1021$ & $35$ &\textcolor{red}{Cancer} \\
&$x_{1}$ & $2025$ & $28$ & \textcolor{blue}{Heart Disease} \\
&$x_{3}$ & $2022$ & $25$ & \textcolor{magenta}{Viral Infection}\\
&$x_{4}$ & $2020$ & $24$ & \textcolor{magenta}{Viral Infection}\\
&$x_{9}$ & $1023$& $31$ & \textcolor{red}{Cancer}\\
\noalign{\smallskip}\hline
\end{tabular}
\end{table}

In the following, we present in detail the major steps of the proposed \textsc{HM-pfsom} algorithm for hybrid microaggregation.
\subsection{The \textsc{HM-pfsom} algorithm for hybrid microaggregation}
The \textsc{HM-pfsom} is a new algorithm for hybrid microaggregation aiming to protect microdata from individual identification, that could be identity or attribute disclosure risk. The \textsc{HM-pfsom} algorithm proceeds by extracting the optimal partition of a given microdata, which will be used to generate the $k$-anonymous microdata. Thus, the obtained microdata can avoid, or at least decrease, the identity disclosure risk. However, the main originality of the \textsc{HM-pfsom} algorithm is that it \textit{redefines} the process of extracting the $k$-partition, by supposing that the cardinality $k$ of each group should not be fixed arbitrary, but it should retain the diversity of confidential attributes. \\
The pseudo-code of the \textsc{HM-pfsom} algorithm is sketched by Algorithm \ref{MicroFSOM}.
\linesnumbered 
\begin{algorithm}[htbp]
 \SetVline
 \setnlskip{-3pt}
 \Entree {
 \begin{itemize}
 \item $X$ : The original microdata 
 \item $k$ : Privacy parameter
 \end{itemize}  
  }
 \Sortie{
  $X'$ : The anonymous microdata
  }
  \Deb{
  Split the microdata $X$ into $c$ disjoint sub-microdata, according to the quasi-identifiers, \textit{i.e.}  $X = Xid_{1} \cup Xid_{2} \cup \ldots \cup Xid_{c} $.\\
 \textcolor{red}{\textit{ /* Apply an hybrid microaggregation process */}}\\
 \ForEach{sub-microdata $Xid_{j} \in \lbrace Xid_{1}, Xid_{2}, \ldots, Xid_{c}  \rbrace$, $\forall j \in \lbrace 1, \ldots, c \rbrace$}{   
 Let $ \bar{x}_{id(j)}$ be the cluster centers of the sub-microdata  $Xid_{j}$.\\
          Split the sub-microdata $Xid_{j}$ into $cs_{j}$ disjoint clusters, according to the confidential attributes.\\  
          Let $Cs = \lbrace C_{1}, C_{2}, \ldots, C_{cs_{j}} \rbrace$ be the set of obtained clusters.\\
		  
\While{ $ \vert C_{i} \vert \geq k, \forall C_{i} \in Cs $}{ 
Extract $x_{r}$ the most distant record to  $  \bar{x}_{id(j)}$.\\
\textcolor{red}{\textit{ /* Form a fixed size group around $x_{r}$*/}}\\
Let $C_{r}$ be the cluster to which belong the data vector  $x_{r}$. \\
Extract from the cluster $C_{r}$ the $(k- 1)$ closest data vector to $x_{r}$\\
Remove the extracted data vector from $ C_{r}$\\
\ForEach{$ C_{i} \in Cs - \lbrace C_{r}\rbrace$}{
Extract the $k$ closest data vector to $x_{r}$\\
Remove the extracted data vector from $ C_{i}$\\
}   
}
Assign the remaining data to their nearest group.\\
 \textcolor{red}{\textit{ /* Generate an anonymous sub-microdata $X'_{j}$ */}}\\
          Within each formed group, replace the values of each quasi-identifier attribute with the average value of the attribute over the group.
          	          } 
          $X' = X'_{1} \cup X'_{2} \cup \ldots \cup   X'_{c} $\\
  }
\caption{The \textsc{HM-pfsom} algorithm}
\label{MicroFSOM}
\end{algorithm}

The \textsc{HM-pfsom} algorithm starts by splitting the original microdata into disjoint sub-microdata (\emph{line $2$}), in such a way that data sharing similar characteristic of quasi-identifiers are gathered in the same sub-microdata. To do so, the \textsc{HM-pfsom} algorithm, relies on fuzzy possibilistic clustering process \citep{Abidi2013}. Thereafter, the partitioning process of the microaggregation can be applied independently on each sub-microdata (\emph{lines $4- 17$}). Accordingly, the risk of gathering dissimilar data in a same group will be eliminated. Thus, the \textsc{HM-pfsom} algorithm avoids the refinement steps used to improve the homogeneity of the final $k$-partition.

To ensure the diversity of confidential attributes within the $k$-partition,  the \textsc{HM-pfsom} algorithm studies the distribution of confidential attributes within each sub-microdata $Xid_{j} \subset X$, $\forall j \in \lbrace 1, \ldots, c \rbrace$. Then, the fixed size groups should be generated in such a way to fulfil the latter distribution. To achieve such purpose, the \textsc{HM-pfsom} algorithm extracts at first,  from each sub-microdata $Xid_{j}$, the $cs_{j}$ disjoint clusters of confidential attributes (\emph{line} $6$). The latter clusters are used thereafter for generating fixed size groups.  Unlike the standard microaggregation methods, the \textsc{HM-pfsom} algorithm imposes a condition on the group size. Each group should gather at least $cs_{j}$ data vectors belonging to different clusters of confidential attributes. To do so, the data vectors of $Xid_{j}$ are distributed into $cs_{j}$ disjoint groups, according to their confidential attributes. These data should be close as possible in terms of quasi-identifiers and distant in terms of confidential attributes. In order to respond to such constraint the \textsc{HM-pfsom} applies an adaptive partitioning process. It computes at first the center of the sub-microdata, noted by $\bar{x}_{id(j)}$ (\emph{line $5$}). Then,  a data vector $x_{r}$ is extracted. The latter corresponds to the most distant data to the centroid $\bar{x}_{id(j)}$ (\emph{line $9$}). Let $C_{r}$ be the cluster to which $x_{r}$ belongs. The \textsc{HM-pfsom} algorithm selects firstly the $k - 1 $ closest data vectors from the clusters $C_{r}$ (\emph{line $12$}). Then, from each cluster $C_{i}$, $\forall i \in \lbrace 1,\ldots, r-1,r+1,\ldots, c_{s} \rbrace$, the $k$ closest data vector to $x_{r}$ are either extracted (\emph{lines $14 - 16$}). Such process is repeated until all fixed size groups, \textit{i.e.} of cardinality $ k \times cs_{j} $, are formed.\\
The remaining data are simply assigned to their closest group. 
Thereby, we can guarantee the diversity of sensitive attributes within each group.\\
Once the input data of the sub-microdata $Xid_{j}$ are partitioned into groups, of cardinality at least $ k \times c_{s}$, the \textsc{HM-pfsom} algorithm generates the anonymous sub-microdata $X'_{j}$, by replacing the data vectors by the centroid to which belong to (\emph{line $19$}).\\
The final anonymous microdata $X'$ is considered as the union of the anonymous sub-microdata (\emph{line $20$}).\\

In the following, we propose to use an illustrative example to highlight the principle of the $k$-partitioning process, adopted by the \textsc{HM-pfsom} algorithm.
\subsubsection*{Illustrative example}
Let  $X_{sub-micro}$, given in Table \ref{submicro}, be a sub-microdata of a given original microdata. Each input data $x_{i} \in X_{sub-micro}$ contains one confidential attribute, \textit{i.e.} \emph{Salary}.
\begin{table}[htbp]
\caption{Example of sub-microdata $X_{sub-micro}$ generated from a given microdata}
\label{submicro}
\centering
\begin{tabular}{cccl}
\hline\noalign{\smallskip}
$x_{i}$ & ZIP code & Age & \textbf{Salary}\\
\noalign{\smallskip}\hline\noalign{\smallskip}
$x_{1}$ & $1011$ & $22$ & $\textbf{500}$\\

$x_{2}$ & $1007$ & $22$ & $\textbf{550}$\\

$x_{3}$ & $1012$ & $23$ & $\textbf{600}$\\

$x_{4}$ & $1009$ & $25$ & $\textbf{1600}$\\

$x_{5}$ & $1010$ & $28$ & $\textbf{1500}$\\

$x_{6}$ & $1011$ & $29$ & $\textbf{1800}$\\

$x_{7}$ & $1013$ & $31$ & $\textbf{2900}$\\

$x_{8}$ & $1010$ & $32$ & $\textbf{3200}$\\

$x_{9}$ & $1008$ & $32$ & $\textbf{3600}$\\

$x_{10}$ & $1010$ & $29$ & $\textbf{1650}$\\

$x_{11}$ & $1009$ & $26$ & $\textbf{1550}$\\

$x_{12}$ & $1011$ & $27$ & $\textbf{1700}$\\

$x_{13}$ & $1008$ & $33$ & $\textbf{3800}$\\
\noalign{\smallskip}\hline
\end{tabular}
\end{table}\\
Before extracting the fixed size groups, the \textsc{HM-pfsom} algorithm starts by applying a clustering process in order to study the distribution of the confidential attribute values.\\
We note that the confidential attribute of the input data are distributed into $3$ clusters, which are respectively illustrated in Tables \ref{low}, \ref{middle} and \ref{high}. Indeed, the set of data $\lbrace x_{1}, x_{2}, x_{3} \rbrace$ are characterized by a salary varying between $500$ and $600$, \textit{i.e.} a low salary values. While the salary attribute of the second set of data vectors $\lbrace x_{4}, x_{5}, x_{6}, x_{10},x_{11},x_{12} \rbrace$ ranges between $1500$ and $1800$, \textit{i.e.} a middle salary values. Whereas the third set of data $\lbrace x_{7}, x_{8}, x_{9}, x_{13} \rbrace$ is characterized by high values of salary, varying between $2900$ and $3800$. \\
\begin{table}[h]
\caption{Cluster $1$: Data vectors with low salary}
\label{low}
\centering
\begin{tabular}{cccl}
\hline\noalign{\smallskip}
$x_{i}$ & ZIP code & Age & \textbf{Salary}\\
\noalign{\smallskip}\hline\noalign{\smallskip}
$x_{1}$ & $1011$ & $22$ & \textcolor{blue}{$\textbf{500}$}\\
$x_{2}$ & $1007$ & $22$ & \textcolor{blue}{$\textbf{550}$}\\
$x_{3}$ & $1012$ & $23$ & \textcolor{blue}{$\textbf{600}$}\\
\noalign{\smallskip}\hline
\end{tabular}
\end{table}\\
\begin{table}[h]
\caption{Cluster $2$: Data vectors with middle salary}
\label{middle}
\centering
\begin{tabular}{cccl}
\hline\noalign{\smallskip}
$x_{i}$ & ZIP code & Age & \textbf{Salary}\\
\noalign{\smallskip}\hline\noalign{\smallskip}
$x_{4}$ & $1009$ & $25$ & \textcolor{red}{$\textbf{1600}$}\\
$x_{5}$ & $1010$ & $28$ & \textcolor{red}{$\textbf{1500}$}\\
$x_{6}$ & $1011$ & $29$ & \textcolor{red}{$\textbf{1800}$}\\
$x_{10}$ & $1010$ & $29$ & \textcolor{red}{$\textbf{1650}$}\\
$x_{11}$ & $1009$ & $26$ & \textcolor{red}{$\textbf{1550}$}\\
$x_{12}$ & $1011$ & $27$ &\textcolor{red}{ $\textbf{1700}$}\\
\noalign{\smallskip}\hline
\end{tabular}
\end{table}\\
\begin{table}[h]
\caption{Cluster $3$: Data vectors with high salary}
\label{high}
\centering
\begin{tabular}{cccl}
\hline\noalign{\smallskip}
$x_{i}$ & ZIP code & Age & \textbf{Salary}\\
\noalign{\smallskip}\hline\noalign{\smallskip}
$x_{7}$ & $1013$ & $31$ & \textcolor{magenta}{$\textbf{2900}$}\\
$x_{8}$ & $1010$ & $32$ & \textcolor{magenta}{$\textbf{3200}$}\\
$x_{9}$ & $1008$ & $32$ & \textcolor{magenta}{$\textbf{3600}$}\\
$x_{13}$ & $1008$ & $33$ & \textcolor{magenta}{$\textbf{3800}$}\\
\noalign{\smallskip}\hline
\end{tabular}
\end{table}

To maintain the latter distribution of the confidential attribute, the \textsc{HM-pfsom} algorithm builds the fixed size groups according to the obtained clusters, by requiring that each group should contain at least $3$ \textit{dissimilar} confidential attribute values. For example, the first fixed size groups $G_{1}$ gathers $3$ data, namely $x_{1}$, $x_{8}$, $x_{10}$ and $x_{12}$, belonging to different clusters. Indeed, the data vector $x_{1}$ has a \textit{low} salary value, while the data vectors $x_{10}$ and $x_{12}$ are characterized, by a \textit{middle} salary. Whereas, the data $x_{8}$ has a \textit{high} salary. The same principle is applied to the other groups, \textit{i.e.} $G_{2}$ and $G_{3}$. Thus, the final $3$-partition should matches to the one given in Table \ref{submicro}.\\
\begin{table}[h]
\caption{The optimal $3$-partition of the sub-microdata $X_{sub-micro}$}
\label{optimalsub}
\centering
\begin{tabular}{ccccl}
\hline\noalign{\smallskip}
&$x_{i}$ & ZIP code & Age & \textbf{Salary}\\
\noalign{\smallskip}\hline\noalign{\smallskip}
\multirow{4}{*}{$ G_{1} $}& $x_{1}$ & $1011$ & $22$ & \textcolor{blue}{$\textbf{500}$}\\

& $x_{12}$ & $1011$ & $27$ & \textcolor{red}{$\textbf{1700}$}\\

& $x_{8}$ & $1010$ & $32$ & \textcolor{magenta}{$\textbf{3200}$}\\

& $x_{10}$ & $1010$ & $29$ & \textcolor{red}{$\textbf{1650}$}\\
\noalign{\smallskip}\hline\noalign{\smallskip}
\multirow{4}{*}{$ G_{2} $} & $x_{2}$ & $1007$ & $22$ &\textcolor{blue}{$\textbf{550}$}\\

& $x_{4}$ & $1009$ & $25$ & \textcolor{red}{$\textbf{1600}$}\\

& $x_{9}$ & $1008$ & $32$ &\textcolor{magenta}{ $\textbf{3600}$}\\

& $x_{11}$ & $1009$ & $26$ & \textcolor{red}{$\textbf{1550}$}\\
\noalign{\smallskip}\hline\noalign{\smallskip}
\multirow{5}{*}{$ G_{3} $} & $x_{3}$ & $1012$ & $23$ & \textcolor{blue}{$\textbf{600}$}\\

& $x_{5}$ & $1010$ & $28$ & \textcolor{red}{$\textbf{1500}$}\\

& $x_{7}$ & $1013$ & $31$ & \textcolor{magenta}{$\textbf{2900}$}\\

& $x_{6}$ & $1011$ & $29$ & \textcolor{red}{$\textbf{1800}$}\\

& $x_{13}$ & $1008$ & $33$ &\textcolor{magenta}{ $\textbf{3800}$}\\
\noalign{\smallskip}\hline
\end{tabular}
\end{table}\\
To achieve such partition, we propose to fix at first the number of groups, denoted by $k_{g}$. Then, the data within each cluster are partitioned into the $k_{g}$ groups. In our example, the number of groups is set equal to $3$, and the data of the clusters \emph{low}, \emph{middle} and \emph{high} salary are distributed into the $3$ groups.

\subsection{Fuzzy possibilistic clustering for hybrid microaggregation}
The main challenge of the \textsc{HM-pfsom} algorithm is to find the suitable set of the sub-microdata contained in the original microdata $X$, \textit{i.e.} $X = \lbrace Xid_{1}, Xid_{2}, \ldots, Xid_{c}\rbrace$. Moreover, for each sub-microdata, it is important to find out its groups of confidential attributes and rightly assess their centres even in noisy surroundings. Clustering is a useful means to achieve such important target. Indeed, clustering aims to discover the unrevealed relationships between data, by splitting a dataset into disjoint clusters. Where each cluster gathers similar data, and dissimilar to those data in other clusters.\\
In this respect, we have proposed in a previous work a fuzzy possibilistic clustering algorithm, called Possibilistic Fuzzy Self Organising Map (\textsc{PFSOM}) \citep{Abidi2013}, able to split a given dataset into $c$ disjoint clusters, via a multi-level process. Where $c$ corresponds to the optimal number of clusters contained in a dataset \citep{Abidi2012}. Each level aims to group similar outputs resulting from the level below. Doing so, the first level is used to form an initial partition of the dataset $X$, by training the data into an initial clusters. The latter clusters are fine-tuned through a hierarchical levels. The role of each level is to build a partition of the outputs of the level below. Such a process produces a hierarchical structure composed of several partitions. To extract the best one, the \textsc{PFSOM} algorithm integrates, during the multi-level process, a validity index, called a \textit{Partition Coefficient and Exponential Separation} (PCAES) index \citep{wu05}. At each level, the algorithm assesses the quality of the partition, by evaluating the compactness and separation of its clusters. The optimal partition $P$ corresponds to the one that have the best validity index. Subsequently, the number of clusters contained in the obtained partition $P$ is equivalent to the optimal number of clusters $c$.\\
At each level, the clustering algorithm \textsc{PFSOM} relies on fuzzy possibilistic learning process, in order to decrease the influence of noisy data. In fact, real datasets are generally characterized by the presence of noisy data and outliers, which can directly influence the obtained data clusters. The \textsc{PFSOM} algorithm integrates both of the concept of typicality and membership values during the clustering process. In fact, to classify a data point, a cluster centroid has to be the closest one to the data point, and this what aims fuzzy clustering by using a probabilistic constraint, \textit{i.e.} membership values \citep{fcm}. In addition, for estimating the centroids, the possibilistic constraint, \textit{i.e.} typicality values, is used for mitigating the undesirable effect of outliers \citep{pcm}. \\
Generally speaking, to split a given dataset $X = \lbrace x_{1}, x_{2}, \ldots, x_{n}\rbrace$  into $c$ clusters, the \textsc{PFSOM} algorithm starts by initializing the $c$ cluster centres. Then, the prototypes of the latter are adjusted during a learning process. This means that, the estimation of the cluster centres is achieved through an iterative process. In each iteration, the prototype of each cluster center $c_{j}$ is updated according to the membership and typicality values of all data to that cluster. Where, the membership value represents the degree to which a given data point $x_{i}$ belongs to a cluster $c_{j}$. Such value is measured according to distances between $x_{i}$ to all cluster centres. However, the typicality of a data point to a given cluster represents  its resemblance to the other data points belonging to the same cluster, \textit{i.e.} internal resemblance. The belonging of a data point $x_{i}$ to a cluster $c_{j}$, depends on the distance from $x_{i}$ to $c_{j}$ relative to the distances of all data to that cluster \citep{fpcm}. The process of updating cluster centres as well as the membership and typicality values is repeated until the stability condition is fulfilled or the predefined
number of iterations is achieved. Then, the clusters are obtained by assigning each data to its nearest center. The clustering process of the \textsc{PFSOM} algorithm  is detailed in \citep{Abidi2013}.

To sum up, the \textsc{HM-pfsom} algorithm relies on two levels of clustering process. On each level, the \textsc{PFSOM} algorithm is used by adapting the distance measure.
In fact, the first level consists in splitting the original microdata $X$ into $c$ disjoint sub-microdata according to their quasi-identifiers, \textit{i.e.} $X = \lbrace Xid_{1}, Xid_{2}, \ldots, Xid_{c}\rbrace$. Thus, the distance used to compute typicality and membership values between a data vector $x_{i}$ and a center $c_{j}$ is defined as follows:
\[
dist(x_{i}, c_{j}) = \sum_{l=1}^{Q} (x_{i}\left[ qi_{h}\right] - c_{j}\left[ qi_{h}\right])
\]
Where $Q$ refers to the number of quasi-identifiers, and $ x\left[ qi_{l}\right] $is the $l^{th}$ quasi-identifier of a given data vector $x$. In the previous example given in Table \ref{submicro}, $Q$ is equal to $2$, while $x_{i}[qi_{1}]$ and $x_{i}[qi_{2}]$ correspond, respectively, to the \emph{ZIP code} and \emph{Age} of the data vector $x_{i}$. The same goes for the center $c_{j}$.\\
On the other hand, the second level of clustering process is used to study the distribution of the confidential attributes within each sub-microdata $Xid_{j}$. It is used to extract the $c_{s}$ disjoint clusters of confidential attributes.
So, the distance used by the PFSOM algorithm is defined as:
\[
dist(x_{i}, c_{j}) = \sum_{p=1}^{S} (x_{i}\left[ s_{p}\right] - c_{j}\left[ s_{p}\right])
\]
Where $S$ refers to the number of confidential attributes, and $ x\left[ s_{p}\right] $
is the $p^{th}$ sensitive attribute of a given data vector $x$. In the example given in Table \ref{submicro}, $S$ is equal to $1$, while $x_{i}[s_{1}]$ corresponds to the \emph{salary} attribute of the $i^{th}$ data vector.\\

\section{Experimental results}
\label{section4}
This section aims to test the validity of the \textsc{HM-pfsom} algorithm for generating an anonymous microdata. We evaluate the performance of our algorithm according to information loss and disclosure risk on well real microdata.\\
In the following we present firstly the manipulated microdata and the measures used to evaluate our algorithm. Then, we discuss the performance values of the \textsc{HM-pfsom} algorithm.
\subsection{Evaluation data}
To evaluate the performance of our proposed \textsc{HM-pfsom} algorithm with the main microaggregation methods, we consider the three real-world microdata, used as benchmarks in prior studies \citep{Ruth2002}, namely:
\begin{itemize}
\item \emph{CENSUS}: This dataset was obtained on July $27$, $2000$ using the Data Extraction System of the U. S. Bureau of the Census. The \emph{CENSUS} dataset contains $1080$ records with $13$ numeric attributes.
\item \emph{EIA}: This dataset was obtained from the U.S. Energy Information Authority. It consists of $4092$ records with $15$ attributes. Since the first two attributes are considered as direct identifiers of the records, we propose to de-identify the dataset by removing the latter attributes. Then, we have not taken into account the attribute $YEAR$, because the value of the latter attribute in the whole dataset is equal to $96$. We also eliminated the categorical attribute $STATE$, given that our proposed algorithm is designed to handle continuous values. To sum up, we used in our experiment only the $11$ remaining attributes.
\item \emph{TARRAGONA}: This real dataset comprises the figures from $834$ companies in the area of Tarragona. This means that, the dataset contains $834$ records with $13$ numeric attributes.
\end{itemize}
Note that, in each dataset, the values of the attributes are well apart, \textit{i.e.} range in different domains. This can distort the clustering results. Thus, we  propose to normalize the manipulated datasets by Min-Max scaling technique. A value $v$ of an attribute is normalized to the value $v'$, by computing the following formula: 
\[
v' = \frac{v-min(v)}{max(v) - min(v)}
\]
Where $min(v)$ and $max(v)$ refer, respectively, to the minimum and the maximum values of the attribute $v$.
\subsection{Evaluation measures}
Let $X$ be an original dataset and $X'$ its anonymous version.
The quality of the microaggregation methods has been evaluated from the perspectives of information loss and disclosure risk, as follows:
\begin{itemize}
\item The information loss (IL) has been quantified by means of the well-known measure which was exposed in \citep{DR}. The IL measure computes the mean variation between the original and the perturbed version of a record $x_{i}$, given by the following formula :
\begin{equation}
IL = \sum_{i=1}^{n}\left(  \frac{1}{q} \times \sum_{j=1}^{q} \frac{ \vert x_{ij} - x'_{ij}\vert }{\sqrt{2}S_{j}}\right) 
\end{equation} 
where $S_{j}$ is the standard deviation of the $j^{th}$ variable in the original data.
Hence, the lower the information loss is, the higher the utility of the anonymous data.
\item The disclosure risk (DR) is quantified by \emph{Distance-Based Record Linkage} (DBRL). This method, introduced in \citep{Pagliuca1999}, consists in computing distances between records in two datasets. The record linkage can be used to find out to what extent anonymous records could be re-identified. In general, for each record in the original dataset $X$, the distance
to every record in the masked dataset $X'$ is computed. Thereafter, the \textit{nearest} and the \textit{second nearest} records in $X'$ are extracted. A record in the anonymous dataset $X'$ is labeled as \textit{linked}, when the nearest record in the original dataset $X$ matches its corresponding original record. A record in the anonymous microdata $X'$ is designed  as linked to \textit{$2^{nd}$ nearest}, if the second nearest record in $X$ turns out to be the corresponding original record. In all other cases, the records are not linked.
\end{itemize}
\subsection{Performance Analysis}
The first part of the experimental evaluation consists in comparing the three main heuristics of microaggregation approach, namely:
\begin{itemize}
\item The univariate microaggregation by individual sorting.
\item The univariate microaggregation based on single axis sorting criteria.
\item The multivariate microaggregation based on diameter method, \textit{i.e.} MDAV.
\end{itemize}
The aim of this comparative study is  to support our choice of using the MDAV algorithm method in the fixed size partitioning process. Then, we discuss the performance of our proposed hybrid microaggregation algorithm.
\subsubsection{Performance Comparaison of univariate and multivariate microaggregation methods}

\begin{figure*}
     \begin{minipage}{0.5\linewidth}
             \begin{center}
                     \includegraphics[width=0.8\textwidth]{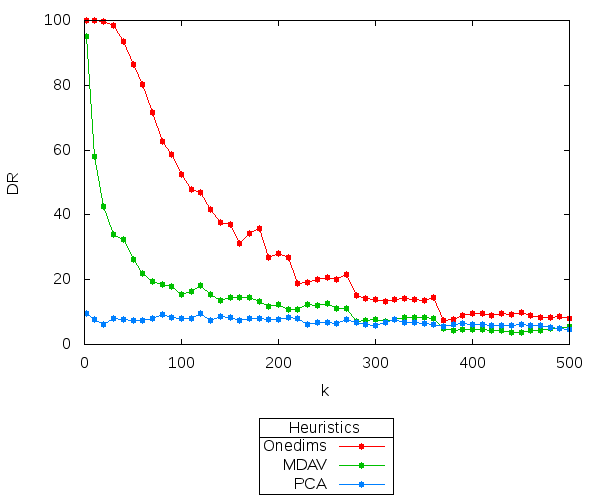}
                     \caption{\small{Comparing the DR of the main microaggregartion heuristics \newline on Census$_{ \vert Qid \vert = 5} $ microdata} }
                     \label{DRCensus}
           
            \end{center}
     \end{minipage}
     \begin{minipage}{0.5\linewidth}
             \begin{center}
                     \includegraphics[width=0.8\textwidth]{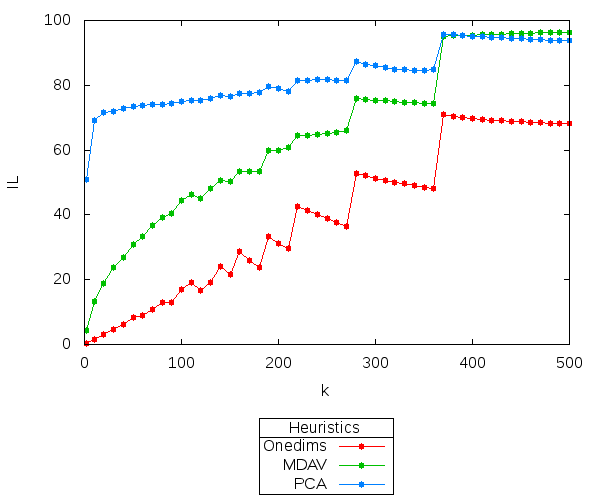}
                     \caption{\small{Comparing the IL of the main microaggregartion heuristics \newline on Census$_{ \vert Qid \vert = 5} $ microdata}}
                     \label{ILCensus}
             \end{center}
     \end{minipage}

\begin{minipage}{0.5\linewidth}
             \begin{center}
                     \includegraphics[width=0.8\textwidth]{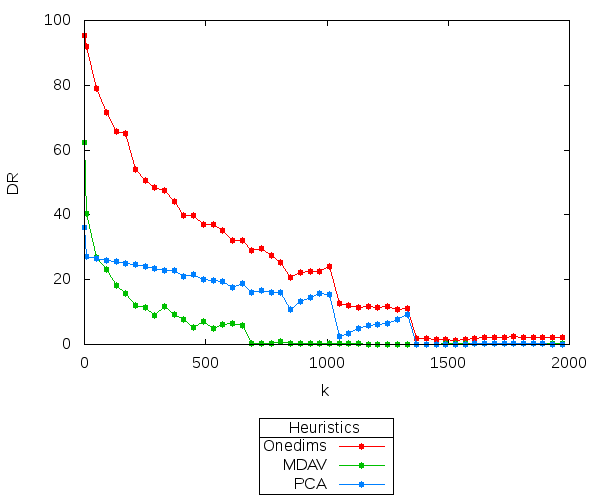}
                     \caption{\small{Comparing the DR of the main microaggregartion heuristics \newline on EIA$_{ \vert Qid \vert = 7} $ microdata} }
                     \label{DREIA}
           
            \end{center}
     \end{minipage}
     \begin{minipage}{0.5\linewidth}
             \begin{center}
                     \includegraphics[width=0.8\textwidth]{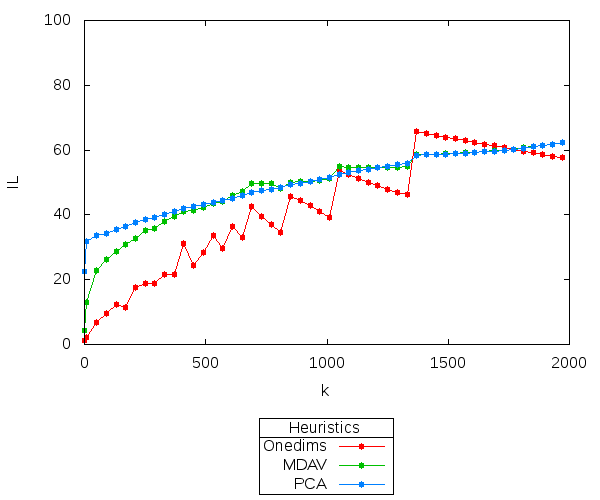}
                     \caption{\small{Comparing the IL of the main microaggregartion heuristics \newline on EIA$_{ \vert Qid \vert = 7} $ microdata}}
                     \label{ILEIA}
             \end{center}
     \end{minipage}

 \begin{minipage}{0.5\linewidth}
             \begin{center}
                     \includegraphics[width=0.8\textwidth]{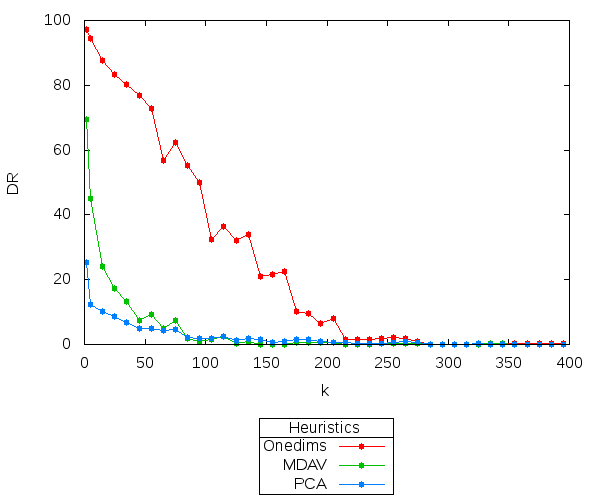}
                     \caption{\small{Comparing the DR of the main microaggregartion heuristics \newline on Tarragona$_{ \vert Qid \vert = 5} $ microdata} }
                     \label{DRTarragona}
           
            \end{center}
     \end{minipage}
     \begin{minipage}{0.5\linewidth}
             \begin{center}
                     \includegraphics[width=0.8\textwidth]{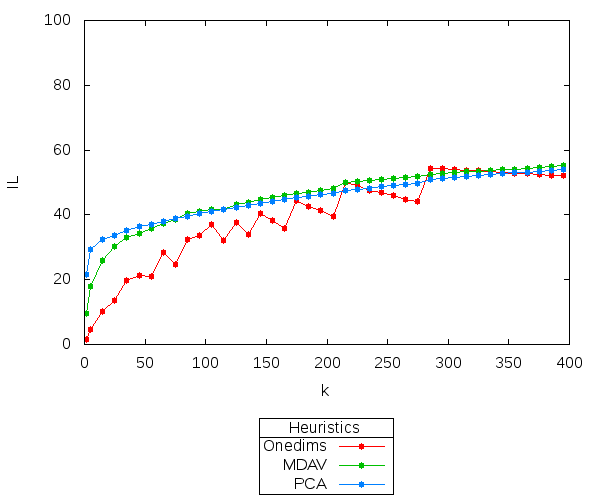}
                     \caption{\small{Comparing the IL of the main microaggregartion heuristics \newline on Tarragona$_{ \vert Qid \vert = 5} $ microdata}}
                     \label{ILTarragona}
             \end{center}
     \end{minipage}      
\end{figure*}

To assess the performance of the microaggregation methods, we used the \emph{R-Package sdcMicro} \citep{Templ2015}. The latter package includes the popular methods of generating protected microdata. In the remainder of this section, we choose to refer the univariate microaggregation heuristic based on individual sorting criteria by ONEDIMS, while the univariate microaggregation based on single axis sorting criteria, by PCA.\\
We evaluate the performance of the three main microaggregation heuristics on the real datasets cited above, by varying the value of the parameter $k$ and the number of the quasi-identifiers. The information loss and disclosure risk measures of the different microaggregation heuristics are given in Figures \ref{DRCensus}, \ref{ILCensus}, \ref{DREIA}, \ref{ILEIA}, \ref{DRTarragona}  and \ref{ILTarragona}.\\
By analyzing the results, we notice that the data utility and the disclosure risk are \textit{inversely} proportional, regardless the microaggregation methods. We note that, for small values of $k$, the information loss is at minimum score, while the disclosure risk reaches its maximum scores. Otherwise, for high values of $k$ the information loss values are increased and the anonymous dataset loses its utility, whereas the risk of records re-identification is minimized. Such a mismatch is confirmed by varying the privacy parameter $k$ on EIA, Tarragona and Census microdata. 
This contradiction can be explained by the fact that for small values of $k$, the records of a given microdata are partitioned into a fixed size groups, \textit{i.e.} of size $k$, gathering similar records. Thus, each record will be close to the centroid of the group to which it belongs to. Thus, replacing the quasi-identifiers of the original records by their centroid, will produce a modified microdata fairly close to the original one. However, this anonymous microdata can not be effective to hide the identity of original records, since an intruder may be able to link the anonymous records with their nearest original ones. On the other hand, by choosing a high value of $k$, the records of the original microdata will be partitioned into a reduced number of fixed size groups. Thereby, records with dissimilar quasi-identifiers will be \textit{forced} to be gathered in a same group. By this way, the centroids of the fixed size groups will be miscalculated. Since the latter will be used to anonymize the microdata, the information loss will be significant, which can avoid the intruder to assume that the centroid assigned to a record is always the nearest one.

By examining the performance of the microaggregation heuristics, namely ONEDIMS, MDAV and PCA, we can note that the multivariate microaggregation, \textit{i.e.} the MDAV method, is best suited than the univariate microaggregation, \textit{i.e.} ONEDIMS and PCA methods. The performance is measured in terms of handling the conflicting principles, \textit{i.e.} maintaining the data utility and avoiding the disclosure risk. Indeed, regardless the value of the parameter $k$, the ONEDIMS method results mostly lead to the lowest information loss, but the highest disclosure risk values. This can be explained by the fact that the univariate microaggregation by individual sorting criteria applies the anonymization process on each attribute. Thus, the anonymous multi-dimensional records can violate the $k$-anonymity property. On the other hand, the univariate microaggregation by PCA sorting criteria method, studies at first the underlying multidimensional \textit{correlation structure} of the data vectors, to  produce  principal  components. Then, the $k$-partition is formed by sorting the training data vectors according to their principal component. Thereafter, groups of successive $k$ records are formed. However, ranking the records by such approach mainly rely  on  the correlation matrix. Such method can be very  useful with data  that  are very  highly correlated. Indeed, the  higher  the  correlation the lower the information loss is. However, not all data are highly correlated.\\
To sum up, we can consider that the multivariate microaggregation is best suited than the univariate microaggregation, in terms of balancing between the two conflicting issues, namely data utility and privacy.\\
Several methods have been proposed to improve the performance of the MDAV method. However, these improvements have been addressed to decrease the information loss. As mentioned above, we think that the optimal $k$-anonymous microdata should not only maintain the information loss while fulfilling the $k$-anonymity property, but also it should avoid attribute disclosure. Thus, we have proposed the \textsc{HM-pfsom} algorithm aiming to achieve the optimal $k$-anonymous microdata, as we have defined. We should remember that our proposed algorithm proceeds through an hybrid manner. It consists in splitting the microdata into disjoint sub-microdata, according to the similarity of the quasi-identifiers. Avoiding thus the risk of gathering records with dissimilar quasi-identifiers in a same group. Thereafter, the microaggregation process can be applied independently on each sub-microdata.

\subsubsection{Performance analysis of hybrid microaggregation}
In the following, we propose to compare the performance of the \textsc{HM-pfsom} algorithm to those of the MDAV algorithm. The performances are evaluated in terms of \textit{i)} homogeneity of sensitive attributes within the fixed size groups, by adapting the well-known \emph{Sum of Squared Errors}; and \textit{ii)}  information loss (IL). \\
Let $P$ be the $k$-partition of the original microdata $X$. 
The homogeneity of sensitive attributes within each group of the $k$-partition $P$ is defined by :
\begin{equation}
SSE_{i} = \sum_{j=1}^{n_{i}} \sum_{p=1}^{s} dist( x_{jp}, c{ip}), \forall i \in \lbrace 1, \ldots, g\rbrace
\end{equation}
Where $SSE_{i}$ measures the distances between the sensitive attributes of the data belonging to a given group $g_{i}$ to its centroid $c_{i}$. The latter centroid is determined by  $\frac{\sum_{j=1}^{n_{i}} \sum_{p=1}^{s}  x_{jp}}{n_{i}}$.
Note that, the lower the value of $SSE_{i}$, the more the sensitive attributes within the $i^{th}$ group are close to their centroid,  which means that the sensitive attributes are similar. Since our aim is to ensure a diversity of sensitive attributes within the fixed size groups, thus high values of $SSE_{i}$ indicate that the sensitive attributes of the $i^{th}$ group are well \textit{separated}.\\
By analyzing the experimental results exposed in Tables \ref{SSECensus} and \ref{SSEEIA1}, we can notice that, regardless the predefined privacy parameter $k$, the \textsc{HM-pfsom} algorithm is able to generate a fixed size partition, where the minimal size of its groups is proportional to the diversity of confidential attributes.\\
Suppose that the Census microdata is composed by a set of $5$ quasi-identifiers and a set of $8$ sensitive attributes. To anonymize the latter microdata, the \textsc{HM-pfsom} algorithm starts by splitting the original microdata into $3$ disjoint sub-microdata, according to the quasi-identifiers. We should mention that we have used the \textsc{PFSOM} to extract the latter sub-microdata.\\
\begin{table*}[htbp]
\begin{center}
\caption{Distribution of the Census microdata}
\label{distcensus}
\begin{tabular}{cll}
\hline\noalign{\smallskip}
Microdata & $\vert sub-microdata \vert $ & $\vert Confidential~classes \vert $ \\
\noalign{\smallskip}\hline\noalign{\smallskip}
\multirow{9}{*}{Census}  & \multirow{3}{*}{$\vert  sub-census_{1} \vert $= 382} & $\vert C_{1} \vert $= 145 \\
 & & $\vert C_{2} \vert  $= 164 \\ 
 &  & $\vert C_{3} \vert $= 73 \\ 
\cline{2-3}
 & \multirow{4}{*}{$\vert sub-census_{2} \vert $= 376} & $\vert C _{1} \vert $= 50 \\
& & $\vert C_{2} \vert  $= 121 \\ 
&  & $\vert C _{3} \vert $=103 \\ 
&  & $\vert C_{4} \vert $=103 \\ 
\cline{2-3}
& \multirow{2}{*}{$\vert sub-census _{3} \vert$= 322} & $\vert C_{1} \vert $= 115 \\
& & $\vert C_{2} \vert $= 207 \\
\noalign{\smallskip}\hline
\end{tabular}

\begin{tablenotes}

      \scriptsize
      \item Table \ref{distcensus} illustrates the $3$ resulted sub-Census microdata, according to the $5$ quasi-identifier attributes. Then, on each sub-Census microdata the distribution of confidential attributes is studied. For example, the data vectors of the first sub-Census$ _{1} $ contains $3$ classes of sensitive attributes. While the second and the third sub-microdata contain respectively $4$ and $2$ classes of sensitive attributes.
    \end{tablenotes}
\end{center}
\end{table*}
\begin{table*}[htbp]
\begin{center}
\caption{Performance comparison on Census microdata} 
\label{SSECensus}
\centering
\begin{small}
\begin{tabular}{clllllllll}
\hline\noalign{\smallskip}
\multirow{3}{*}{ sub-microdata} & \multirow{3}{*}{k} &\multicolumn{2}{c}{minimal} &\multicolumn{2}{c}{\multirow{2}{*}{\# Groups} } & \multicolumn{2}{c}{\multirow{2}{*}{min $SSE _{j, 1\leq j \leq cs}$} } & \multicolumn{2}{c}{\multirow{2}{*}{IL}}\\

& &\multicolumn{2}{c}{group size} &\multicolumn{1}{c}{} &  &\multicolumn{1}{c}{} &  & \multicolumn{1}{c}{}&\\ 
\noalign{\smallskip} \cline{3-10}\noalign{\smallskip}
&  &\textsc{HM-pfsom} & \textsc{mdav} & \textsc{HM-pfsom}  & \textsc{mdav}& \textsc{HM-pfsom}  & \textsc{mdav} &\textsc{HM-pfsom} & \textsc{mdav}\\

\noalign{\smallskip}\hline\noalign{\smallskip}
\multirow{6}{*}{sub-Census$ _{1} $}& 1 & 4 & 1 & 73 & 382 & 4.46 & 0.00 & 22.75 & 0.00\\

& 5 & 20 & 5 & 14 & 76 & 29.76 & 3.72 & 31.07 & 15.90\\
& 10 & 40 & 10 & 7 & 38 & 67.92 & 10.90 & 32.96 & 20.75\\
& 15 & 60 & 15 & 4 & 25 & 133.33 & 14.31 & 35.05 & 23.11\\
& 20 & 82 & 20 & 3 & 19 & 149.11 & 23.63 & 36.84 & 24.82\\
& 25 & 100 & 25 & 2 & 15 & 262.48 & 35.56 & 40.31 & 26.03\\
\noalign{\smallskip}\hline\noalign{\smallskip}
\multirow{5}{*}{sub-Census$ _{2} $}&1 & 7 & 1 & 50 & 370 & 4.52 & 0.00 & 25.54 & 0.00\\

&5 & 35 & 5 & 10 & 75 & 27.26 & 2.66 & 31.24 & 13.98\\
&10 & 70 & 10 & 5 & 37 & 63.07 & 8.21 & 36.22 & 18.69\\
&15 & 105 & 15 & 3 & 25 & 104.27 & 12.44 & 37.05 & 20.64\\
&20 & 140 & 20 & 2 & 18 & 184.53 & 16.29 & 41.17 & 22.63\\
\noalign{\smallskip}\hline\noalign{\smallskip}
\multirow{6}{*}{sub-Census$ _{3} $} &1 & 2 & 1 & 115 & 322 & 2.58 & 0.00 & 17.90 & 0.00\\
&5 & 10 & 5 & 23 & 64 & 13.07 & 6.30 & 25.89 & 16.30\\
&10 & 20 & 10 & 11 & 32 & 48.65 & 10.84 & 29.90 & 20.84\\
&20 & 40 & 20 & 5 & 16 & 151.60 & 30.60 & 33.37 & 24.74\\
&30 & 60 & 30 & 3 & 10 & 267.17 & 61.67 & 37.15 & 27.88\\
&40 & 80 & 40 & 2 & 8 & 445.00 & 72.94 & 40.87 & 30.52\\
\noalign{\smallskip}\hline
\end{tabular}
\end{small}
\begin{tablenotes}
\scriptsize
 \item  The performances are evaluated in terms of information loss (IL) and diversity of sensitive attributes within the fixed size groups ($SSE_{i}$). The higher the value of $SSE_{i}$, the more the sensitive attributes within the $i^{th}$ group are dissimilar. Thus, the attribute disclosure risk is low. 
\end{tablenotes}
\end{center}
\end{table*}

\begin{table*}[htbp]
\begin{small}
\begin{center}
\caption{Splitting the EIA microdata into disjoint sub-microdata according to the $7$ quasi-identifiers.}
\label{disteia}
\centering
\begin{tabular}{cll}
\hline\noalign{\smallskip}
Microdata & $\vert sub-microdata \vert $ & $\vert Confidential~classes \vert $ \\
\noalign{\smallskip}\hline\noalign{\smallskip}
\multirow{12}{*}{EIA} & \multirow{3}{*}{ $\vert sub-EIA_{1} \vert$ = 464} & $\vert C_{1}\vert$ = 238 \\
 & &  $\vert C_{2}\vert$ = 78 \\
& & $\vert C_{3}\vert$ = 148\\
\cline{2-3}
 & \multirow{3}{*}{ $\vert sub-EIA_{2} \vert$ = 1195} & $\vert C_{1}\vert$ = 828 \\
 & &   $\vert C_{2}\vert$ = 233 \\
 & & $\vert C_{3}\vert$ = 134 \\
 \cline{2-3}
 & \multirow{2}{*}{ $\vert sub-EIA_{3} \vert$ = 106} & $\vert C_{1}\vert$ = 37 \\
& &   $\vert C_{2}\vert$ = 69\\
 \cline{2-3}
& \multirow{2}{*}{ $\vert sub-EIA_{4} \vert$ = 1191} & $\vert C_{1}\vert$ = 273 \\
& &   $\vert C_{2}\vert$ = 918\\
 \cline{2-3}
 & \multirow{2}{*}{ $\vert sub-EIA_{5} \vert$ = 1136} & $\vert C_{1}\vert$ =259 \\
& &   $\vert C_{2}\vert$ = 877\\
\noalign{\smallskip}\hline
\end{tabular}
\end{center}
\end{small}
\begin{tablenotes}
\scriptsize
\item  Table \ref{disteia} illustrates the $5$ resulted sub-EIA microdata, according to the $7$ quasi-identifiers of the data records. The third column shows the distribution of confidential attributes on each sub-EIA. 
\end{tablenotes}
\end{table*}
\begin{table*}[htbp]
\begin{center}
\caption{Performance comparison on EIA microdata}
\label{SSEEIA1}
\begin{small}
\begin{tabular}{clllllllll}
\hline\noalign{\smallskip}
\multirow{3}{*}{ sub-microdata} & \multirow{3}{*}{k} &\multicolumn{2}{c}{minimal} &\multicolumn{2}{c}{\multirow{2}{*}{\# Groups} } & \multicolumn{2}{c}{\multirow{2}{*}{min $SSE _{j, 1\leq j \leq cs}$} } & \multicolumn{2}{c}{\multirow{2}{*}{IL}}\\

& &\multicolumn{2}{c}{group size} &\multicolumn{1}{c}{} &  &\multicolumn{1}{c}{} &  & \multicolumn{1}{c}{}&\\ 
\noalign{\smallskip} \cline{3-10}\noalign{\smallskip}
&  &\textsc{HM-pfsom} & \textsc{mdav} & \textsc{HM-pfsom}  & \textsc{mdav}& \textsc{HM-pfsom}  & \textsc{mdav} &\textsc{HM-pfsom} & \textsc{mdav}\\

\noalign{\smallskip}\hline\noalign{\smallskip}
\multirow{6}{*}{sub-EIA$ _{1} $}& 1 & 5 & 1 & 78 & 464 & 4.89 & 0.00 & 21.21 & 0.00\\
& 5 & 25 & 5 & 15 & 92 & 34.61 & 4.60 & 29.09 & 12.97\\
& 10 & 50 & 10 & 7 & 46 & 88.69 & 8.74 & 33.44 & 17.51\\
& 20 & 100 & 20 & 3 & 23 & 275.50 & 15.85 & 41.78 & 21.51\\
& 30 & 150 & 30 & 2 & 15 & 457.89 & 42.64 & 46.45 & 23.84\\
\noalign{\smallskip}\hline\noalign{\smallskip}
\multirow{6}{*}{sub-EIA$ _{2} $} &1 & 8 & 1 & 134 & 1195 & 3.15 & 0.00 & 50.47 & 0.00\\
& 5 & 40 & 5 & 26 & 239 & 23.23 & 260 & 50.46 & 50.75\\
& 10 & 80 & 10 & 13 & 11 & 62.14 & 5.73 & 50.46 & 50.46\\
& 20 & 160 & 20 & 6 & 59 & 130.53 & 18.55 & 50.45 & 50.48 \\
& 30 & 240 & 30 & 4 & 39 & 262.95 & 27.02 & 50.45 & 50.46\\
& 40 & 320 & 40 & 3 & 29 & 432.04 & 36.02 & 50.43 & 40.47\\
\noalign{\smallskip}\hline\noalign{\smallskip}
\multirow{4}{*}{sub-EIA$ _{3} $}& 1 & 2& 1 & 37&106 & 8.54 & 0.00 & 26.07 & 0.00\\
& 5 & 10 &5 & 7 &21 & 58.26 &12.13 & 32.88 &17.95 \\
& 10 & 20 & 10& 3& 10 & 134.27 &29.50 & 40.44& 25.49\\
& 15 & 30 &15 & 2 &7 & 211.60& 59.06 & 40.02 & 27.55\\
\noalign{\smallskip}\hline\noalign{\smallskip}
\multirow{9}{*}{sub-EIA$ _{4} $}& 1 & 4 & 1 & 273 & 1191 & 1.08 & 0.00 & 15.48 & 0.00\\
&5 & 20 & 5 & 54 & 238 & 7.33 & 0.10 & 18.52 & 5.18\\
&10 & 40 & 10 & 27 & 119 & 13.91 & 0.55 & 20.94 & 7.47\\
&30 & 120 & 30 & 9 & 39 & 72.60 & 2.68 & 25.25 & 11.08\\
&50 & 200 & 50 & 5 & 23 & 146.44 & 4.65 & 30.57 & 13.13\\
&100 & 400 & 100 & 2 & 11 & 474.85 & 10.74 & 46.75 & 17.22\\
\noalign{\smallskip}\hline\noalign{\smallskip}
\multirow{9}{*}{sub-EIA$ _{5} $} &1 & 4 & 1 & 259 & 1136 & 1.09 & 0.00 & 14.69 & 0.00\\
& 5 & 20 & 5 & 51 & 227 & 6.67 & 0.06 & 14.69 & 14.80\\
& 10 & 40 & 10 & 25 & 113 & 15.68 & 0.61 & 14.64 & 14.76\\
& 30 & 120 & 30 & 8 & 37 & 65.44 & 2.99 & 14.56 & 14.76\\
& 50 & 200 & 50 & 5 & 22 & 137.34 & 4.28 & 14.50 & 14.72\\
& 100 & 400 & 100 & 2 & 11 & 465.27 & 15.11 & 14.14 & 14.67\\
\noalign{\smallskip}\hline
\end{tabular}
\end{small}
\begin{tablenotes}

      \scriptsize
      \item  The performances are evaluated in terms of information loss (IL) and diversity of sensitive attributes within the fixed size groups ($SSE_{i}$). The higher the value of $SSE_{i}$ is, the more the sensitive attributes within the $i^{th}$ group are dissimilar.

    \end{tablenotes}
\end{center}
\end{table*}
Table \ref{distcensus} illustrates the $3$ resulted sub-Census microdata, which contain respectively 382, 376 and 322 data records. On each sub-microdata, the \textsc{HM-pfsom} algorithm studies in a second phase the distribution of the confidential attributes. For example, such as mentioned in Table \ref{distcensus}, the data vectors of the first sub-microdata of Census, \textit{i.e.} sub-Census$ _{1} $, are in turn partitioned into $3$ classes of sensitive attributes. While the second and the third sub-microdata contain respectively $4$ and $2$ classes of confidential attributes. These classes are then used to maintain the diversity within the obtained partition. For example, by setting the parameter $k$ to the minimal value, \textit{i.e.} $k=2$, the obtained partition from sub-Census$ _{1} $ is formed by $36$ groups having a minimal size equal to $8$. Indeed, each group contain at least $2$ records from the three classes of confidential attributes. In fact, we note that the minimal homogeneity of sensitive attributes $SSE_{i}$ within the obtained groups is equal $8.41$. While the minimal $SSE_{i}$ values obtained by the MDAV algorithm is equal to $0.9 \simeq 0$. This would mean that the $k$-partitioning process of the MDAV algorithm can group data within a same group having a \textit{highly} similar sensitive attributes. These data are then exposed to the attribute attack. By increasing the value of the parameter $k$, the \textsc{HM-pfsom} algorithm is able to maintain the diversity of sensitive attributes within the resulted partition. Indeed, by setting the parameter $k$ to $25$ the minimal $SSE_{i}$ value obtained by \textsc{HM-pfsom} algorithm is equal to $268.48$. While that of the MDAV algorithm is equal to $35.56$. However, as expected, ensuring a diversity of confidential attributes within the $k$-partition may affect the data utility. In fact, the performance of the information loss obtained by the MDAV algorithm are better than those of the \textsc{HM-pfsom} algorithm. 
Indeed, on the three sub-microdata of Census microdata, by increasing the value of the privacy parameter, the information loss obtained by \textsc{HM-pfsom} algorithm is increased from 22.75 to 40.31 for sub-Census$ _{1} $; from 25.54 to 41.17 for sub-Census$ _{2} $; and from 17.90 to 40.87 for sub-Census$ _{3} $. Whereas, the information loss obtained by the MDAV algorithm increases from 8.84 to 26.03 for sub-Census$ _{1} $, from 8.01 to 26.03  for sub-Census$ _{2} $ and from 12.57 to 30.52 for sub-Census$ _{3} $.
The detailed observations are given in Table \ref{SSECensus}. \\
These low performances in terms of data utility, is also true for the sub-microdata tables of EIA microdata, which are illustrated in Table \ref{SSEEIA1} (page \pageref{SSEEIA1}). This can be explained by the fact that data with similar quasi-identifiers can be assigned to different groups, since they have a close sensitive attributes. By contrast, this can avoid the attribute disclosure risk.
\section{Conclusion}
\label{section5}
In this paper, we have introduced a new algorithm for multivariate microaggregation \textsc{HM-pfsom}, based on fuzzy possibilistic clustering to generate the optimal partition. The latter is used to generate an anonymous microdata. Hence, the \textsc{HM-pfsom} algorithm covers three main goals, namely: \textit{i)} Preventing the identity disclosure risk by requiring that the generated partition should fulfill the $k$-anonymity property; \textit{ii)} Preventing the attribute disclosure by ensuring the diversity of confidential attributes within each fixed size group of the generated partition; \textit{iii)} Maintaining the data utility of the anonymous microdata by increasing the homogeneity of the obtained partition.\\
The \textsc{HM-pfsom} algorithm operates through an hybrid manner. Its main idea consists in splitting the original dataset into disjoint sub-datasets, in such a way that data within the same sub-dataset must be similar to some extend, also they should be dissimilar to those data in other sub-datasets. Such process enables to avoid the refinement phases used to fine tune the $k$-anonymous partition, since the partitioning process is performed only on similar data. Indeed, \textsc{HM-pfsom} applies the $k$-partitioning process independently on each sub-dataset. On privacy side, the \textsc{HM-pfsom} approach proposes to study the distribution of confidential attributes within each sub-dataset. Then, according to the latter distribution, the privacy parameter $k$ is determined, in such a way to preserve the diversity of confidential attributes within the anonymized microdata.

Our main interest in a future work lies in Privacy-Preserving Data Sharing in the Smart City. In fact, smart city is a vision proposed by many governments to integrate information and communication technology (ICT) solutions into the critical infrastructures of their cities and society with the goal of improving the quality of life of their citizens \citep{Curry2016}. Smart city applications comprise a number of diverse areas, like smart card services for easy authentication and payment on the go, smart resource management of water or electricity, smart mobility applications that improve traffic efficiency and reduce $CO_{2}$ emissions \citep{Novotny2014}. The effectiveness of these and other smart city applications heavily relies on data collection, interconnectivity, and pervasiveness. Smart city applications are increasingly relying on personally identifiable data. In fact, people's data are key elements in order to design effective and smart policies and services for citizens. Such type of data reflects the daily activities of the people living, working, and visiting the city, \textit{e.g.} monitoring tourist foot traffic, or home energy usage, or homelessness \citep{Zoonen2016}. The more connected a city the more it will generate a steady stream of data from and about its citizens. However, connected smart city devices raise concerns about individuals' privacy, autonomy, freedom of choice, and potential discrimination by institutions.  Thus, privacy  is  a  key  concern  in  the  facet  of smart  cities.  Thereby, develop a new framework for exploring people's specific privacy concerns in smart cities is considered as an obvious prospect.
The framework should hypothesize if and how smart city technologies and urban big data
produce privacy concerns among the people in these cities, such as inhabitants, workers, visitors, and otherwise.

\bibliographystyle{spbasic}
\bibliography{Bibliography}
\end{document}